\documentclass[11pt]{article}
\usepackage[margin=1in]{geometry}
\usepackage{natbib}
\usepackage{hyperref}
\usepackage{enumerate}
\usepackage{graphicx} 
\usepackage{todonotes}
\usepackage{booktabs}
\usepackage{amsthm}
\newtheorem*{researchquestion}{Research Question}
\newtheorem{designgoal}{Design Goal}

\title{On Large Language Models as Data Sources for Policy Deliberation on Climate Change and Sustainability}
\author{Rachel Bina \\
 rfbina@gmail.com \and Kha Luong \\
 khaluong@wharton.upenn.edu
 \and Shrey Mehta \\ shreym@sas.upenn.edu  \and Daphne Pang \\
 daphne.y.pang@gmail.com \and
 Mingjun (Kristen) Xie \\
 mingjunx@seas.upenn.edu 
  \and Christine Chou \\ chou@gms.ndhu.edu.tw \and  Steven O. Kimbrough \\
 sok@upenn.edu
 \and (and the Penn Climate Decisions Lab)}

\date{\today}

\begin{document}

\maketitle

\begin{quote}
    {\it Democracy begins with conversation.}

\flushright    ---John Dewey at his 90\textsuperscript{th} birthday party\footnote{Lamont, C. (Ed.). (1959). Dialog on John Dewey. Horizon Press. Page 88.
}

\flushleft {\it  Only that which has no history can be defined.}

\flushright ---Frederick Nietzsche {\it On the Genealogy of Morals}

\flushleft {\it The definition of the problem, rather than its solution, will be the scarce resource in the future.}
\flushright
                        ---Esther Dyson (1990s)

\end{quote}

\begin{abstract}
We pose the research question,
{\it 
    Can LLMs provide credible evaluation scores, suitable for constructing starter MCDM models that support commencing deliberation regarding climate and sustainability policies?}
In this exploratory study we
\begin{enumerate}[i.]
\item Identify a number of interesting policy alternatives that are actively considered by local governments in the United States (and indeed around the world).
\item Identify a number of quality-of-life indicators as apt evaluation criteria for these policies.
\item Use GPT-4 to obtain 
evaluation scores for the policies on multiple criteria. 
\item Use the TOPSIS MCDM method to rank the policies based on the obtained evaluation scores.
\item Evaluate the quality and validity of the resulting 
ensemble of scores
by comparing the TOPSIS-based policy rankings with those obtained by an informed assessment exercise.
\end{enumerate}
We find that GPT-4 is in rough agreement with the policy rankings of our informed assessment exercise. Hence, we conclude (always provisionally and assuming a modest level of vetting) that GPT-4 can be used as a credible input, even starting point, for subsequent deliberation processes on climate and sustainability policies.
\end{abstract}


\tableofcontents

\section{Context\label{sec_context}}


This study occurs in the context of a larger project that is directed at, among other things, two design goals.  The first of these goals is about providing deliberation  tools for supporting decision making on  climate and sustainability policies.

\begin{designgoal}[tools]\label{design_goal_tools}
    Construct tools to support deliberation on climate and sustainability policies. These tools may come in any of several forms, including knowledge of methods and practices, as well as databases, document corpora, and software.  The list of potential forms is open-ended. 
\end{designgoal}

The second motivating design goal is more specific. It constitutes a way of achieving instances that meet the first goal. The second goal is about MCDM (multi-criteria decision-making) models. These constitute a broad class of modeling methods that are designed and intended to facilitate decision-making in the presence of two or more conflicting goals and objectives. This kind of situation is rife in everyday life, for example, when we make purchasing decisions that have to trade off cost and quality. Policy decisions in climate change and sustainability are more complex, typically having more than two goals or evaluation criteria that have to be considered. The goals are typically in some degree of conflict and themselves complex.

MCDM models, while long established and used, are under-employed. This is in significant part due to the fact that developing a model requires specialized expertise, which may not be available or affordable in particular cases, as well as rather intense and laborious feedback from decision makers, who are busy and often skeptical of yet-to-be-seen tools \citep{keeney_value-focused_1992,keeney_decisions_1993}. This occasions our second design goal.

\begin{designgoal}[starter MCDM models]\label{goal_mcdm_models}
    Deliver to climate and sustainability policy stakeholders MCDM models (``starter MCDM models'') that are {\it prima facie}  acceptable, comprehensible, and suitable for follow on deliberation, revision, and evolution. These initial models should require minimal input from stakeholders, while affording revision in light of new information and stakeholder views when available.
\end{designgoal}

 We foresee being able to achieve Design Goal \ref{goal_mcdm_models} and thereby being able to afford processes in which technical analysts develop {\it prima facie} valid (alias ``not stupid'')  starter MCDM models that may be used in public deliberations. Under extreme time constraints, such models would be usable without substantial revision for decision making. (This is not to say that the model's assessments are simply accepted; rather, the assessments are subject to deliberation with the information available.) In the more usual case, the model is revised and evolved in light of new information and public comments.  In this way, an ever-improving, presumably valid knowledge base contributes to ongoing discussions and is available whenever decisions have to be made.

 With these goals before us, we hypothesize that large language models (LLMs), such as GPT-4, can be instrumental in creating starter MCDM models that meet Design Goal \ref{goal_mcdm_models}.

\begin{researchquestion}\label{research_question}
    Can LLMs 
    provide credible evaluation scores, suitable for constructing starter MCDM models that support commencing deliberation regarding climate and sustainability policies?
\end{researchquestion}

This is all quite aspirational. Certainly, no single study can meet the design goals and dispositively answer the research question. This paper reports on an exploratory study that bears on the research question in the context of the two design goals.

The upshot of our study is to give  a positive answer to the research question. We turn now to essential framing and setup for what is to follow.
\vskip 9 pt
\centerline{* * *}
\vskip 9 pt

 Table \ref{acs_template}  is a schema or template for what we shall alternatively call an ACS (\underline{a}lternatives, evaluation \underline{c}riteria, \underline{scores}) or $P$ (\underline{p}erformance) table. Such tables are essential data objects when comparing multiple alternatives (the $a_i$ policy alternatives) on multiple dimensions of value (the $c_j$ evaluation criteria). The entries $s_{i,j}$ in the table represent the evaluation scores for their associated $a_i$ and $c_j$ alternatives and evaluation criteria: $s_{i,j}$ is the evaluation score for alternative $a_i$ on criterion $c_j$. 


\begin{table}[h]
\centering
\begin{tabular}{r|cccccc}
Policy & \multicolumn{6}{c}{Evaluation Criteria} \\ 
Alternatives & $c_1$ & $c_2$ & \ldots & $c_j$ & \ldots & $c_n$ \\ \hline
$a_1$ & $s_{1,1}$ & $s_{1,2}$ & \ldots & 
$s_{1,j}$ & \ldots & $s_{1,n}$\\
$a_2$ & $s_{2,1}$ & $s_{2,2}$ & \ldots & 
$s_{2,j}$ & \ldots & $s_{2,n}$\\
\vdots & \vdots & \vdots & \vdots & \vdots & \vdots & \vdots \\
$a_i$ & $s_{i,1}$ & $s_{i,2}$ & \ldots & 
$s_{i,j}$ & \ldots & $s_{i,n}$\\
\vdots & \vdots & \vdots & \vdots & \vdots & \vdots & \vdots \\
$a_m$ & $s_{m,1}$ & $s_{m,2}$ & \ldots & 
$s_{m,j}$ & \ldots & $s_{m,n}$\\
\end{tabular}
\caption{General form---schema or template---for an ACS (alias $P$) table
.}
\label{acs_template}
\end{table}
Given a complete performance table, $P$, decision makers have something definite and useful to work with in deliberating  which alternatives to select for implementation. Of course, additional information may be useful and desired. Even so, an ACS ($P$) table is a fundamental requirement and starting point for serious deliberation. We emphasize \emph{starting point} because modeling and the deliberation it supports should above all be seen as dynamic processes, halted only by practical considerations. Our concern here is with methods for arriving at  data and models that are useful because they can contribute to supervening processes. Not final results, but results positioned for getting better results.\footnote{And that can do well enough if necessary.}

This raises the question of how ACS tables of usefully high quality can and should be constructed to support decision making in specific contexts. We are especially concerned in this study with climate and sustainability policies and with prospects for reducing the workload of scoring $P$ tables.  Specifically, this study addresses the paper's   Research Question, page \pageref{research_question}, \S\ref{sec_context}.

In what follows, we 
\begin{enumerate}[a.]
\item Identify a number of interesting policy alternatives that are actively considered by local governments in the United States (and indeed around the world).
\item Identify a number of quality-of-life indicators as apt evaluation criteria for these policies.
\item Use ChatGPT-4 to obtain  scores ($s_{i,j}$ values in Table \ref{acs_template}) to complete an ACS ($P$) table.
\item Evaluate the quality and validity of the resulting $P$ table by comparing its policy rankings with those obtained by an informed assessment exercise using human judgments.
\end{enumerate}

The report is organized into several sections. \S\ref{sec_background}, ``Background,''  extends this introductory section and provides a broader and deeper account of our multiple criteria decision making (MCDM) approach to climate and sustainability policy deliberation. This section may be skipped on first reading.

\S\ref{sec_lit_review}, ``Review of the literature,'' focuses on recent work exploring the validity of responses obtained from large language models, especially GPT. Frequent findings that large language models ``hallucinate''---give responses in terms of non--existent entities or are wildly wrong in other ways---are certainly worrisome in the context of our research question. While ultimately we are content to treat large language models as black boxes for our purpose of scoring ACS tables, awareness of systematic failures of these models is  certainly relevant to any overall answer to our research question.

\S\ref{sec_gpt_instructions}, ``GPT-4 Instructions,'' describes the process and format used to pose the queries to GPT-4 that produced the scores in our GPT-4 ACS table. This section is supplemented by two appendices, \ref{sec_ref_policies} and \ref{sec_ref_criteria}.
Appendix \ref{sec_ref_policies}, ``Reference Policy Set for the Study,''
presents and describes the collection of climate and sustainability policies---in terms of the ACS table, the alternatives---we considered in our study.
Appendix \ref{sec_ref_criteria}, ``Reference Evaluation Criteria for the Study,'' presents and describes nine quality-of-life evaluation criteria used for scoring the alternative policies.

\S\ref{sec_gpt4_acs}, ``GPT-4 ACS table,'' presents and discusses the ACS table resulting from the queries posed regarding performance (on quality of life, mitigation, adaption) to GPT-4.

\S\ref{sec_mcdm_results}, ``MCDM results,'' presents and discusses application of MCDM modeling to the GPT-4 ACS table. This results in a preference ranking of the policy alternatives under consideration.

\S\ref{sec_human_scoring},  ``Comparison with human scoring,'' works with an ACS table covering a superset of the policies in the GPT ACS table. We employed an \emph{informed assessment} process in which members of the Climate Decisions Lab were given background information on the the policy alternatives and the evaluation criteria, then asked to report a score for the corresponding cell in the ACS table. Each cell was scored by multiple people. These scores were averaged to produce the final reported scores. This resulted in which we call the \emph{informed assessment ACS table} or IA ACS table. We applied MCDM modeling as we did for the GPT-4 ACS table to obtain a (modeled) preference ranking of the policy alternatives. We then discuss the ranking and compare it to the ranking obtained from the GPT-4 ACS table. We find substantial agreement between the two rankings. This leads us to conclude---always provisionally---that LLMs can be used for starter or baseline models for policy deliberation, presumably as a prelude to collecting and incorporating additional data from more credible sources.

\S\ref{sec_conclusion}, ``Conclusion,'' concludes the paper with a summary and discussion of the significance of our findings and methods, and suggests directions for future investigation.

\section{Background\label{sec_background}}

The ``triple planetary crises'' of pollution, climate change,  and biodiversity loss, present a decision-making challenge---locally as well as globally---that is fraught and enormously complex \citep{environment_making_2021}. 
There are well more than 100 triple planetary crisis plans, programs, and policies (alias ``climate and sustainability policies,'' ``policies'') known to be available to local governments (counties and municipalities) to address these crises.\footnote{More than 1500 are identified in \citep{stechemesser_climate_2024}. 
However, these are tokens of policy types. The paper categorizes these instances into 2--3 score types of policies.   Moreover, these are national-level policies, whereas we focus on locally administered and managed policies.
}\textsuperscript{,}\footnote{We have assembled more than 100 policies in the spreadsheet file LocalPoliciesDatabase.xlsx, which may be found at \url{https://tinyurl.com/yyfydfnp}.} Each policy (type or kind) in turn is associated with multiple program configurations and subject to many distinct local jurisdictions. Complicating matters, in no place can the available policies all be implemented at one time. At best, a portfolio of policies can be funded and managed at a particular time and place and for a particular planning cycle. Which policies should be implemented and in what order? What does implementation actually look like in detail? Who is involved in the decision-making process? Fundamentally, how are policies to be compared and evaluated for purposes of decision making? These questions present enduring challenges, if only because of the computational complexity involved.\footnote{With only ten policies under consideration there are about 16 million possible portfolios to consider.} This last question---posing the \emph{policy evaluation problem} for climate and sustainability policies---is the main focus of the present study.

There are broadly three kinds of justifications used for policy evaluation and adoption. The first is financial advantage. If there is profit to be made or money to be saved with the policy, if a cost-benefit analysis is favorable, this is \emph{ceteris paribus} justification for the policy. Climate adaptation policies are often justified in this way. For example, fire protection policies and building code updates may warrant their expense by prospects of cost avoidance in the future.\footnote{As we see by recent events in Los Angeles. A similar point can be made about storm water management policies, both for wet and dry climates.} They serve as a sort of insurance protection against storm damage. Altruism is the second kind of justification. If the cost-benefit analysis of a policy is judged not to favor its adoption, the policy may be undertaken nevertheless because of its overall resulting social good. For example, a policy that reduces greenhouse gas (GHG) emissions or that materially assists the disadvantaged but is costly may be undertaken in the name of climate change on broadly moral grounds. Some individuals behave this way by installing uneconomic rooftop solar PV, for example. They  act nobly, altruistically, and prosocially. 

Altruism, however, is a vulnerable and unreliable basis for public policy; it is unlikely to be sustainable. Hence, there remains the problem of finding justification, if justification can  be had, for seemingly uneconomic but still valuable policies.

The third kind of justification occurs when the policy in question fails or comes out uncertain on a strict cost--benefit basis but is judged to have sufficient co-benefits to overcome a cost--benefit shortcoming \citep{boyd_mitigation_2022,creutzig_demand-side_2022,dagnachew_climate_2022,finn_much_2023,karlsson_climate_2020,sharifi_co-benefits_2021}. An example of such a policy might be a ban on single-use plastic bags (yielding co-benefits of reduced litter in the streets and reduced landfill material) or a ban on gasoline-powered leaf blowers (yielding co-benefits of cleaner air, reduced ecological damage, and reduced noise). 

The present study centers upon this third kind of co-benefit-based warrant for climate and sustainability policies, and within it on factors that contribute to quality-of-life (alias well-being).
Certain aspects of climate and sustainability plans, such as costs, emissions, degree of protection, etc., are technical by nature and best evaluated by climate scientists, city planners, engineers, and area experts in general, or obtained through careful assessment of the literature. Other aspects, particularly those focused on how a policy could impact  well-being and quality of life in the community, perhaps should not be assessed by policy makers except through consultation with representatives of the general public. These \emph{well-being} aspects of climate and sustainability policies are our main, but not exclusive, concern in this study.  

Our study presumes that climate and sustainability policies are usefully---or even necessarily---compared across multiple evaluation criteria and that these criteria are inevitably in conflict, even limiting our attention to quality-of-life and well-being criteria. No policy is always and everywhere the best, and trade-offs are inevitable.

Given this, we further assume the value of constructing multiple criteria decision making (MCDM) models for comparative assessment of climate and sustainability policies. Briefly, MCDM models,\footnote{Some example sources among  many others: \citep{alvarez_multiple-criteria_2021,arrow_social_1986,ishizaka_multi-criteria_2013,kleinmuntz_resource_2007,wu_validation_2018}.} whichever of many particular methods are employed, rely fundamentally on a common data structure. We call this the ACS (\underline{a}ltnatives, evaluation \underline{c}riteria, and \underline{s}cores) table, or alternatively the $P$ (\underline{p}erformance) table. To construct an MCDM model one must at minimum
\begin{enumerate}[A.]
    \item Identify a consideration set of alternatives, $a_i \in \mathcal{A}$, among which choice is to be made.
    \setcounter{enumi}{2}
    \item Identify multiple evaluation  criteria, $c_j \in \mathcal{C}$,  with which to evaluate the alternatives.
    \setcounter{enumi}{18}
    \item Evaluate each alternative $i$ on every evaluation criterion $j$ to obtain a score, $s_{i,j}$.
\end{enumerate}
Once an ACS table is constructed, any of several dozen MCDM methods can be applied to gain insight for decision making.  Step C, scoring, while challenging, is perhaps the step most plausibly aided by automation and information retrieval on large corpora. The immediate purpose of this study can be framed as addressing the following research question.

\begin{researchquestion}[Repeated from \S\ref{sec_context}, page \pageref{research_question}]
    Can LLMs 
    provide credible evaluation scores, suitable for constructing starter MCDM models that support commencing deliberation regarding climate and sustainability policies?
\end{researchquestion}

Our study is necessarily exploratory. Having framed the policy evaluation problem as usefully addressed as an MCDM (multiple criteria decision making) problem, in what follows we:

\begin{enumerate}
    \item Select a consideration set of policies that are salient for local decision makers in America and, we believe, well beyond (Appendix \ref{sec_ref_policies}).
    \item Identify nine quality-of-life indicators plus two additional criteria pertinent to climate and sustainability policies (Appendices \ref{sec_ref_criteria} and \ref{sec_qol_indicators}).
    \item Use ChatGPT to assess each policy in the policy consideration set on each evaluation criterion.
    \item Apply the TOPSIS MCDM method to the resulting ACS table thereby ranking the policies on the assessed evaluation criteria.
    \item Discuss critically how good and how credible the ChatGPT answers are by comparing the results to the results of an informed assessment exercise performed by members of the Climate Decisions Lab.
\end{enumerate}

\subsection{Further points arising}

\begin{enumerate}
    \item The ACS/$P$ table structure is simple and easily understood: 
    \begin{enumerate}[i.]
    \item Rows correspond to policy \underline{a}lternatives. $A$ refers to the collection of alternatives, $a_i$ to a member of that collection.
    \item Columns correspond to evaluation criteria. $C$ refers to the collection of evaluation \underline{c}riteria, $c_j$ to a member of that collection.
    \item ACS table entries represent numerical \underline{s}cores for the corresponding alternatives and evaluation criteria. The collection of such scores is denoted $S$ and $s_{i,j}$ is an element of $S$. Ancillary tables or data structures may record non-numeric information pertaining to scores, such as textual comments.
    \end{enumerate}
    \item The ACS/$P$ table format does not formally assume a linear or additive overall evaluation function on the alternatives. This is, however, an assumption often made when comparing scored alternatives with each other.\footnote{Given our purpose of overviewing an approach, we defer the technical questions of additivity, preferential independence, etc. to another venue.}     \item The $s_{i,j}$'s may be obtained from  three main sources:
    \begin{enumerate}
        \item Published credible accounts, e.g., from peer reviewed literature. These sources are thin for local policies and well-being.
        \item Subject matter experts. These sources are also thin for local policies and well-being.
        \item Stakeholder deliberations and informed assessment exercises.
        The present study relies on informed assessment for the
$s_{i,j}$ scores in the ACS/$P$ table. The informed assessors were well-informed (but not expert) members of the research team, principally graduate and undergraduate students and a few faculty. Informed assessors  are in principle always available. What  has to be kept in mind is that they may be greatly mistaken or subtly biased. They may do their best, but that will not be perfect. This is an inevitable source of biased and inaccurate data, but it is the best available.  We proceed on the basis of a \textit{\textbf{draft and revise}} deliberation philosophy.
 
    \end{enumerate}

\end{enumerate}

Our $P$ table for the informed assessment exercise is given in Table \ref{tbl_p_table} on page \pageref{tbl_p_table}.\footnote{The ACS table and the $P$ table are equivalent. When we include the row labels with the $P$ table we call it the ACS table.} The scores were arrived at by a group process of \emph{informed assessment}. Assessors received instruction on the policies, and assessments were averaged across two or more assessors after discussion.

\section{Review of the Literature\label{sec_lit_review}}

We propose to explore whether large language models, and GPT-4 in particular, can effectively and reliably supply scores in ACS tables for climate and sustainability policies.
It has not escaped notice that the `safety' of GPT responses is often problematic, in that false and even fanciful (``hallucinogenic'') responses often occur. This section reviews a sampling of the literature that investigates this issue, which is obviously important in the context of our proposed use of GPT.

\subsection{Huang et al., 2024}

\cite{huang_survey_2024} is a comprehensive survey of the state-of-the-art of hallucination research on large language models (LLMs), such as LLaMA, Claude, Gemini, and GPT-4. The latter being the chosen tool for our study.

\begin{figure}[h]
    \centering
    \includegraphics[width=1\linewidth]{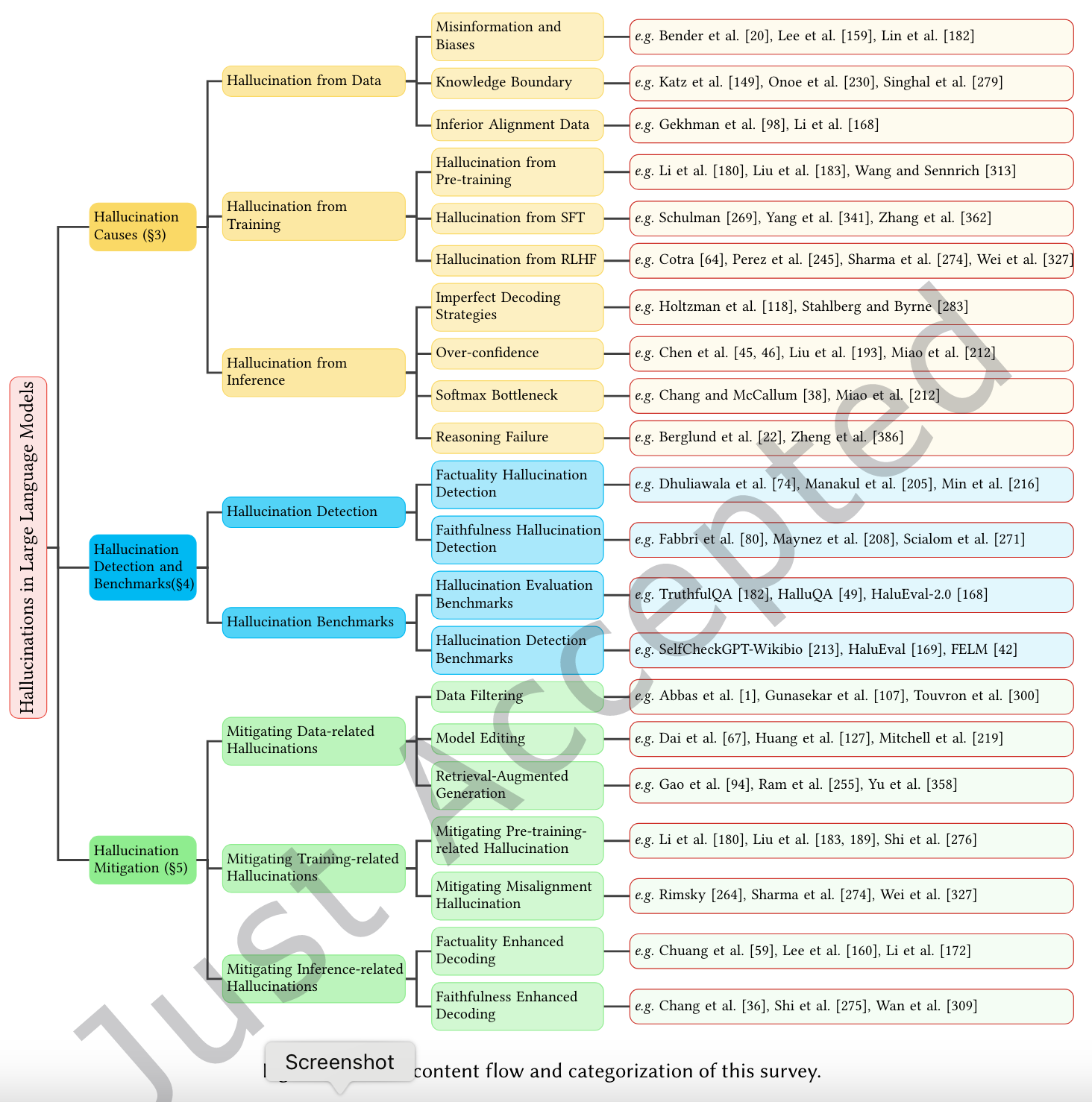}
    \caption{The main content flow and categorization of the survey \cite{huang_survey_2024}.}
    \label{fig_huangetal_fig1}
\end{figure}

\begin{quote}
    [H]allucinations in conventional natural language generation (NLG) tasks have been extensively studied \ldots, with hallucinations defined as generated content that is either nonsensical or unfaithful to the provided source content. These hallucinations are categorized into two types: intrinsic hallucination, where the generated output contradicts the source content, and extrinsic hallucination, where the generated output cannot be verified from the source.
\end{quote}
The study categorizes
\begin{quote}
hallucination into two primary types: \emph{factuality hallucination} and \emph{faithfulness hallucination}. Factuality hallucination emphasizes the discrepancy between generated content and verifiable real-world facts, typically manifesting as factual inconsistencies. Conversely, faithfulness hallucination captures the divergence of generated content from user input or the lack of self-consistency within the generated content. This category is further subdivided into instruction inconsistency, where the content deviates from the user’s original instruction; context inconsistency, highlighting discrepancies from the provided context; and logical inconsistency, highlighting internal contradictions within the content. This categorization refines our understanding of hallucinations in LLMs, aligning it closely with their contemporary usage.    
\end{quote}
The extent and thoroughness of the survey is considerable. See Figure \ref{fig_huangetal_fig1}. For present purposes, the main findings may be summarized as follows.
\begin{enumerate}
    \item Hallucination in LLMs is indeed a pervasive and troubling problem.
    \item Factuality hallucinations are more readily detectable and remediable than faithfulness hallucinations. Note: our proposed use of LLMs, for filling in ACS tables, is exposed to both kinds of hallucination.
    \item A variety of detection and remediation techniques have been proposed and have been found to be effective to some extent in particular circumstances. There is, however, as yet no normative and definite (exactly correct) general procedure for doing this.
    \item The picture that emerges is one of indefinite probing of LLM responses, drawing from an evolving pool of methods, to be halted by pragmatic considerations.
    \item In reinforcement of the previous point, we observe that there is very little discussion of Type I versus Type II errors. Presumably, reducing hallucinations (Type I errors) also reduces the scope of returned claims (Type II errors).  But see page 24 of \cite{huang_survey_2024} on the ``inherent trade-off between diversity and context attribution.'' Type I--Type II dilemmas are likely inherent in any scheme to reduce hallucination.  This too militates in favor of an evolutionary, indefinite probing approach to using results from LLMs, at least in our context.
\end{enumerate}

\newpage

\subsection{OpenAI, 2024}

\begin{figure}[h]
    \centering
    \includegraphics[width=1\linewidth]{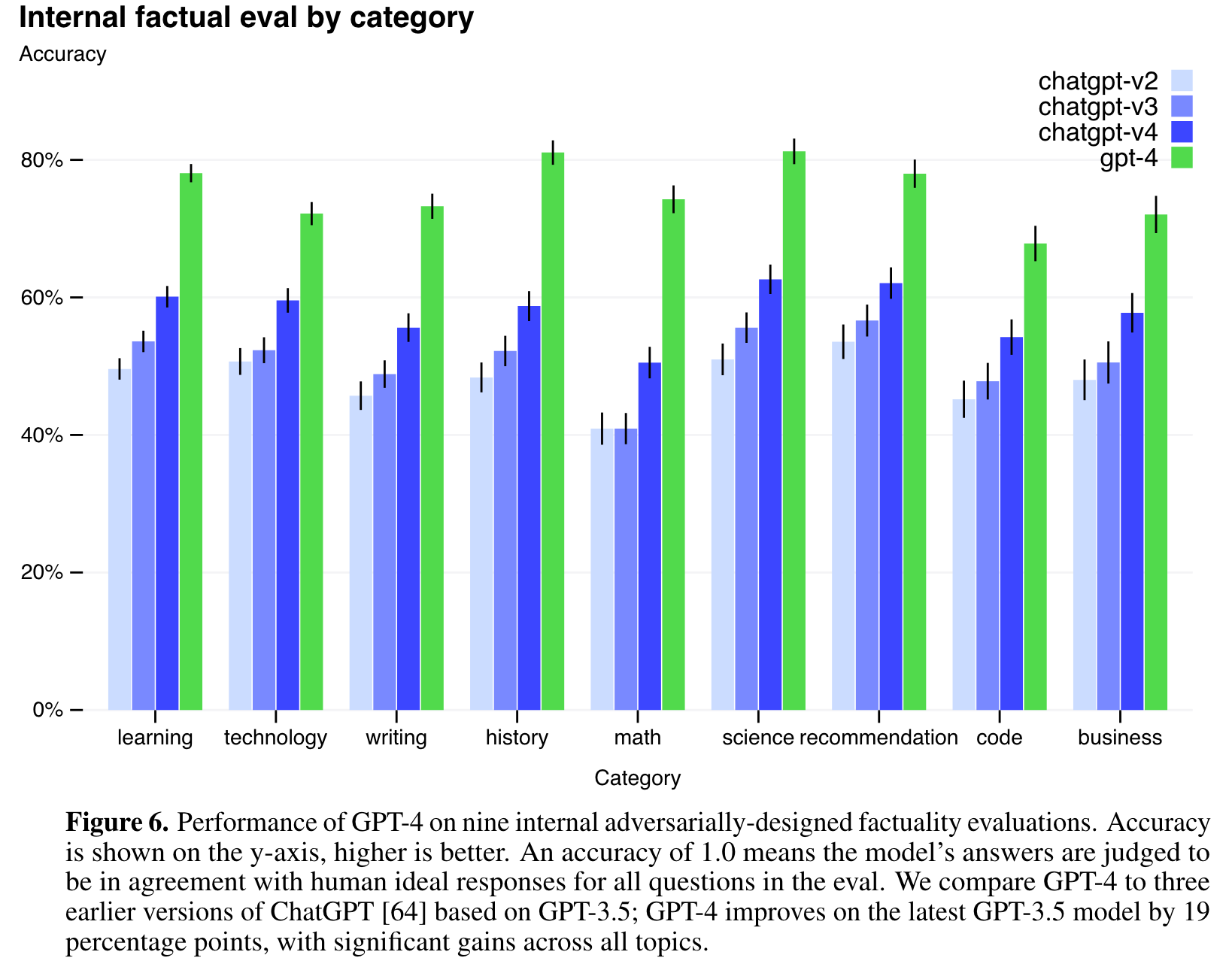}
    \caption{From \citep{openai_gpt-4_2024}.}
    \label{fig_gpt-4}
\end{figure}

\cite{openai_gpt-4_2024} is a recent extensive study of hallucination in GPT-4 by a large group at OpenAI. The study documents the considerable accomplishments of GPT-4, for example its scoring the in the top 10\% on bar exams, among many other challenging tests. These are summarized in Table 1 of \citep{openai_gpt-4_2024}. The study goes on to examine in detail failure rates in different contexts. These results are reproduced from the paper in our Figure
\ref{fig_gpt-4}. Although none of the nine categories examined are identical with ours, it would be prudent and sensible to presume similar performances.

The written summary/comment on these results is surely well-taken:
\begin{quote}
    Despite its capabilities, GPT-4 has similar limitations as earlier GPT models. Most importantly, it still is not fully reliable (it “hallucinates” facts and makes reasoning errors). Great care should be taken when using language model outputs, particularly in high-stakes contexts, with the exact protocol (such as human review, grounding with additional context, or avoiding high-stakes uses altogether) matching the needs of specific applications.
    \citep[page 10]{openai_gpt-4_2024}
\end{quote}

\subsection{Also of note}

\subsubsection{Chelli et al.}

Chelli, M., Descamps, J., Lavoué, V., Trojani, C., Azar, M., Deckert, M., Raynier, J.-L., Clowez, G., Boileau, P., \& Ruetsch-Chelli, C. (2024). Hallucination Rates and Reference Accuracy of ChatGPT and Bard for Systematic Reviews: Comparative Analysis. Journal of Medical Internet Research, 26(1), e53164. 
\url{https://www.jmir.org/2024/1/e53164}

Comparative Analysis
\begin{enumerate}[a.]
\item	Objectives: The aim of the study is to assess the performance of LLMs such as ChatGPT and Bard (subsequently rebranded Gemini) to produce references in the context of scientific writing.
\item	Results: In total, 11 systematic reviews across 4 fields yielded 33 prompts to LLMs (3 LLMs×11 reviews), with 471 references analyzed. Precision rates for GPT-3.5, GPT-4, and Bard were 9.4\% (13/139), 13.4\% (16/119), and 0\% (0/104) respectively (P<.001). Recall rates were 11.9\% (13/109) for GPT-3.5 and 13.7\% (15/109) for GPT-4, with Bard failing to retrieve any relevant papers (P<.001). Hallucination rates stood at 39.6\% (55/139) for GPT-3.5, 28.6\% (34/119) for GPT-4, and 91.4\% (95/104) for Bard (P$<$.001). 
\end{enumerate}
These dispiriting findings stand in contrast to the comparatively upbeat results reported in \citep{openai_gpt-4_2024}. Unfortunately, the tasks here are likely more similar to our proposed uses than are the tasks in \citep{openai_gpt-4_2024}, which probe for established facts.

\subsubsection{Li et al., HaluEval, 2023}

Li, J., Cheng, X., Zhao, W. X., Nie, J.-Y., \& Wen, J.-R. (2023). HaluEval: A Large-Scale Hallucination Evaluation Benchmark for Large Language Models (arXiv:2305.11747). arXiv. 
\newline \url{https://doi.org/10.48550/arXiv.2305.11747}

The paper introduces HaluEval, a large-scale benchmark for evaluating hallucination recognition in Large Language Models (LLMs). The benchmark consists of 35,000 samples, including 5,000 human-annotated ChatGPT responses to general queries and 30,000 automatically generated task-specific examples for question answering, knowledge-grounded dialogue, and text summarization. The authors propose a two-stage framework for generating hallucinated samples: sampling-then-filtering. They find that ChatGPT generates hallucinated content in about 19.5\% of responses, primarily by fabricating unverifiable information. Experiments show that existing LLMs struggle to recognize hallucinations, with even ChatGPT achieving only 62.59\% accuracy in question answering.The study also explores strategies to improve hallucination recognition, such as providing external knowledge and adding reasoning steps. 

\subsubsection{Dahl et al.}

Dahl, M., Magesh, V., Suzgun, M., \& Ho, D. E. (2024). Large Legal Fictions: Profiling Legal Hallucinations in Large Language Models. Journal of Legal Analysis, 16(1), 64–93. \newline https://doi.org/10.1093/jla/laae003
arXiv:2401.01301 [cs] http://arxiv.org/abs/2401.01301

\begin{enumerate}[a.]
\item 	PDF Access: https://arxiv.org/pdf/2401.01301 

\item	Objective and Results: This paper develops a typology of legal hallucinations, providing a conceptual framework for future research in this area. Second, it finds that legal hallucinations are alarmingly prevalent, occurring between 58\% of the time with ChatGPT 4 and 88\% with Llama 2, when these models are asked specific, verifiable questions about random federal court cases. Third, it illustrates that LLMs often fail to correct a user's incorrect legal assumptions in a contra-factual question setup. Fourth, this paper provides evidence that LLMs cannot always predict, or do not always know, when they are producing legal hallucinations. 
\end{enumerate}
\subsection{Upshot}

\begin{enumerate}
\item Hallucination problems for LLMs are amply documented and well-established.
\item This  summary and comment on these problems for GPT-4 is surely well-taken in general for LLMs:
\begin{quote}
    Despite its capabilities, GPT-4 has similar limitations as earlier GPT models. Most importantly, it still is not fully reliable (it “hallucinates” facts and makes reasoning errors). Great care should be taken when using language model outputs, particularly in high-stakes contexts, with the exact protocol (such as human review, grounding with additional context, or avoiding high-stakes uses altogether) matching the needs of specific applications.
    \citep[page 10]{openai_gpt-4_2024}
\end{quote}
\item Progress has been made in mitigation techniques. These hold promise of materially increasing confidence in results, but  on a case-by-case basis.
\item Together, these findings warrant provisional use of LLMs for scoring in ACS tables, in conjunction with vetting results and continued, evolving mitigation.
\end{enumerate}

\subsection{Aside: ChatGPT-o1's epistemic status}

For what it is worth:

Request posed:
\begin{quote}
    Please comment on the epistemic status of your replies to questions. How warranted are they? How reliable? You will surely want to qualify your answer by type of question, and by subject matter. (2024-09-27)
\end{quote}

Answered with
\begin{quote}
    My responses aim to be accurate and reliable based on the information available up to my last update in October 2023. The warrant and reliability of my replies can vary depending on the type of question and subject matter:

- **Factual Questions:** For well-established facts in areas like history, science, and general knowledge, my answers are generally dependable. However, I may lack information on events or developments that occurred after my last update.

- **Specialized or Technical Subjects:** In fields that require deep expertise—such as advanced sciences, niche technologies, or specialized academic topics—my knowledge is as accurate as the data I've been trained on. While I strive for precision, there may be gaps or outdated information due to the rapidly evolving nature of these fields.

- **Subjective or Opinion-Based Questions:** For queries involving opinions, ethics, or personal perspectives, I provide balanced viewpoints based on a wide range of sources. These responses are meant to inform rather than persuade and should not be taken as definitive answers.

- **Creative or Open-Ended Queries:** When addressing creative tasks like writing prompts, brainstorming ideas, or exploring hypotheticals, my replies are designed to inspire and engage. These are generated based on patterns and examples in the data and may not reflect unique or groundbreaking ideas.

Overall, while I strive to offer accurate and helpful information across various topics, it's important to consult authoritative sources or professionals for critical decisions or specialized knowledge.
\end{quote}
A fair assessment, if not as specific,  detailed, and incisive as one might hope.

\section{GPT-4 instructions\label{sec_gpt_instructions}}

In querying GPT and LLMs generally, we wish to ask for an evaluation of a specific policy on a specific evaluation criterion.  For this, we need a standard name and description of the policy, as well as a name and description of the evaluation criterion. Then we assemble them with a template into a standardized query for ChatGPT.

Here is the template used for all queries with minor modifications:

\begin{quote}
Consider first a sustainability policy of \textbf{$<$policy name$>$}, practiced at the county or municipal of government. {\bf $<$policy description.$>$}

Consider second, a policy evaluation criterion:  {\bf$<$evaluation criterion name$>$}. {\bf$<$evaluation criterion description.$>$}

How would you rate the \textbf{$<$policy name$>$} policy on a 1 to 10 scale for its capacity to do well on {\bf$<$evaluation criterion name$>$} considerations?
\end{quote}

See 
the ``Policy Descriptions'' chapter of the supplemental materials for the `official' policy names and descriptions used for prompting ChatGPT with this template.

See Appendix \ref{sec_qol_indicators} and 
the ``Quality of Life'' chapter
of the supplemental material for the `official' quality-of-life names and descriptions used for prompting ChatGPT with this template. See 
the ``Criteria in Addition to QoL'' chapter
of the supplemental for the `official' names and descriptions of other evaluation criteria to be used for prompting ChatGPT. These are for adaptation and mitigation.

See the ``Basic query template for LLM querying'' section of the Supplementary Material for further information.

Records of the queries posed to and responses given by GPT-4 can be found in Climate Decisions Lab (sok)/Research Projects/LLMs and Policy MCDM/.


\newpage
\section{GPT-4 ACS table\label{sec_gpt4_acs}}

\begin{figure}[h]
        \centering
        \includegraphics[width=1\linewidth]{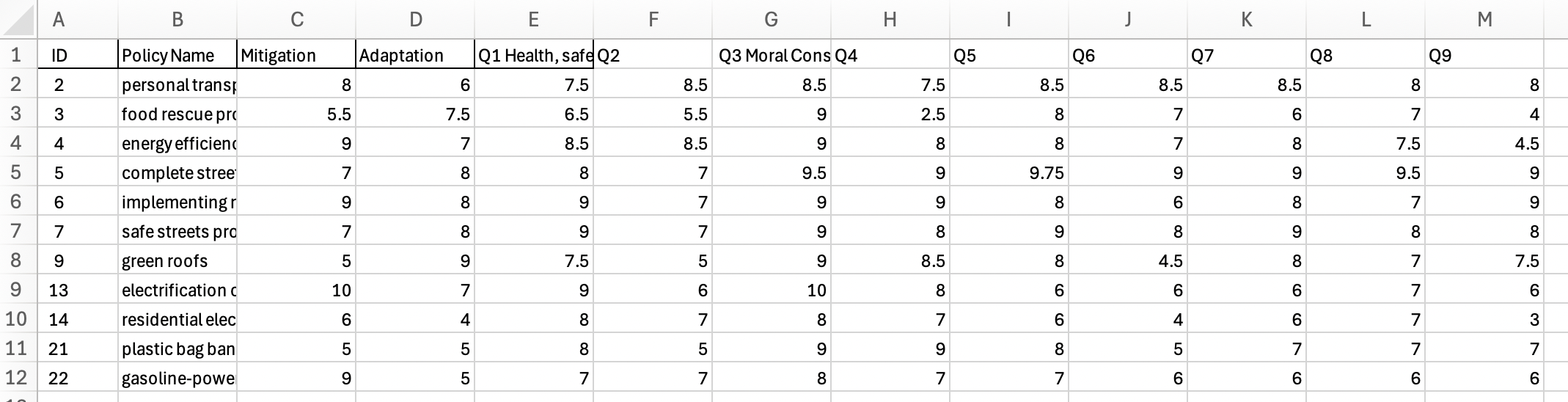}
        \caption{Thirteen polices and their evaluations. File: CurrentDatabase.xlsx, at \url{https://tinyurl.com/26azw7pp}.}
        \label{fig_chatgpt_acs}
    \end{figure}

The ACS table shown in Figure \ref{fig_chatgpt_acs} is the result of consolidating the ChatGPT-4 queries constructed to fit the template of the previous section.

How valid or reliable this ACS table is can only be determined after sufficient probing, \emph{pace} the discussion of mitigation in \S \ref{sec_lit_review}. We note in this regard that we retained all of the ChatGPT responses from which these scores were obtained and the responses uniformly contained information that explained (at least to a degree) why the numerical score was given as it was.  For example, we posed the following question in seeking a score for cell D12. 
\begin{quote}
    Leaf blower bans

Consider first a sustainability policy of Gasoline-Powered Leaf Blowers Ban, practiced at the county or municipal of government. Some municipalities are banning or phasing out gas-powered lawn mowers and leaf blowers in favor of electric alternatives to reduce noise, pollution, and greenhouse gas emissions.  Consider second, a policy evaluation criterion: adaptation to climate change. Adaptation in the context of climate change refers to the process of adjusting systems (natural, social, or economic) to minimize the negative impacts of climate changes while maximizing potential benefits. It involves understanding the current and projected impacts of climate change and developing strategies to cope with them, ensuring resilience and sustainability.
 How would you rate the Gasoline-Powered Leaf Blowers Ban policy on a 1 to 10 scale for its capacity to do well on adaptation considerations?

\end{quote}

Here is ChatGPT-4's response:
\begin{quote}
    To evaluate the Gasoline-Powered Leaf Blowers Ban on its capacity to address adaptation to climate change, it is essential to analyze its direct and indirect contributions to resilience and sustainability in the face of climate change. Here's how it aligns:
    
Strengths

Reduction of Local Air Pollution: Gasoline-powered leaf blowers contribute to smog, particulate matter, and other pollutants that exacerbate respiratory and cardiovascular issues, especially in heat-stressed urban areas. Reducing this pollution supports public health adaptation.
Mitigation Synergy: While primarily a mitigation policy (reducing greenhouse gas emissions), transitioning to electric alternatives fosters cleaner air and reduces heat-trapping emissions, indirectly aiding adaptation strategies like improved air quality in urban heat islands.
Encouragement of Sustainable Practices: Switching to electric alternatives aligns with broader sustainability goals, fostering community awareness and behavior change that supports adaptive mindsets.
Weaknesses
Indirect Link to Adaptation: The policy's primary focus is on mitigation rather than direct adaptation. While it reduces emissions and pollution, it doesn't explicitly address climate resilience, such as urban cooling, stormwater management, or ecosystem restoration.
Equity Concerns: Electric alternatives may impose higher upfront costs, potentially creating barriers for lower-income households or small landscapers. Unequal access to resources can hinder overall community resilience.
Energy Source Dependence: If electricity for alternatives comes from fossil fuels, the benefits might be partially offset, reducing the effectiveness of the policy in fostering a fully adaptive and sustainable energy system.

Rating: 6/10

The ban contributes to adaptation indirectly through improved air quality and reduced emissions, which are important for building community resilience to climate impacts. However, its primary focus on mitigation and potential equity challenges limit its full effectiveness as a climate adaptation policy.

\end{quote}
This is upon review a prima facie sensible and valid answer. Compare it to ChatGPT's answer for climate mitigation:

\begin{quote}
    To rate the Gasoline-Powered Leaf Blowers Ban policy on its capacity to contribute to greenhouse gas mitigation, we need to evaluate its impact in terms of emission reduction compared to other available strategies. Here's a structured breakdown:
    
Strengths:

Reduction in Fossil Fuel Use: Gasoline-powered leaf blowers emit greenhouse gases, including CO2 and unburned hydrocarbons. Banning them directly cuts these emissions.
\vskip 9 pt
Transition to Electric Alternatives: Electric leaf blowers are typically powered by a grid increasingly transitioning to renewable energy sources, resulting in lower lifecycle emissions.
Localized Impact: Although small-scale, this policy directly impacts individual behavior and equipment usage within the municipality, contributing to community-wide reductions.

Limitations:

Scope of Emission Reduction: Gas-powered leaf blowers are a relatively minor source of emissions compared to sectors like transportation, energy production, and agriculture.

Dependent on Electricity Source: The mitigation impact depends on the local energy grid's reliance on renewable energy versus fossil fuels. If the grid is coal or gas-heavy, the net mitigation benefit could be lower.

Behavioral Changes Required: The effectiveness depends on the adoption of electric alternatives and proper disposal of old gas-powered equipment to avoid additional environmental impact.

Relative Impact:

While this policy addresses a specific source of emissions and aligns with broader sustainability goals, its mitigation potential is modest compared to other strategies such as transitioning to renewable energy or improving energy efficiency in buildings and transportation. However, it serves as an important symbolic and incremental step toward broader climate action.

Rating:

On a 1 to 10 scale, where 1 is negligible impact and 10 is transformative climate mitigation, the Gasoline-Powered Leaf Blowers Ban policy could be rated at 4 to 5. This reflects its moderate capacity to mitigate emissions, acknowledging both its localized benefits and its relatively limited scale of impact.
\end{quote}

With the understanding that GPT is grading on an absolute scale of some sort, we find that on balance the generated explanations redound to the credibility of the scores produced. The scores can be taken as \emph{prima facie} valid, subject revision under necessary subsequent probing.

\section{MCDM results\label{sec_mcdm_results}}

\begin{figure}
    \centering
    \includegraphics[width=1\linewidth]{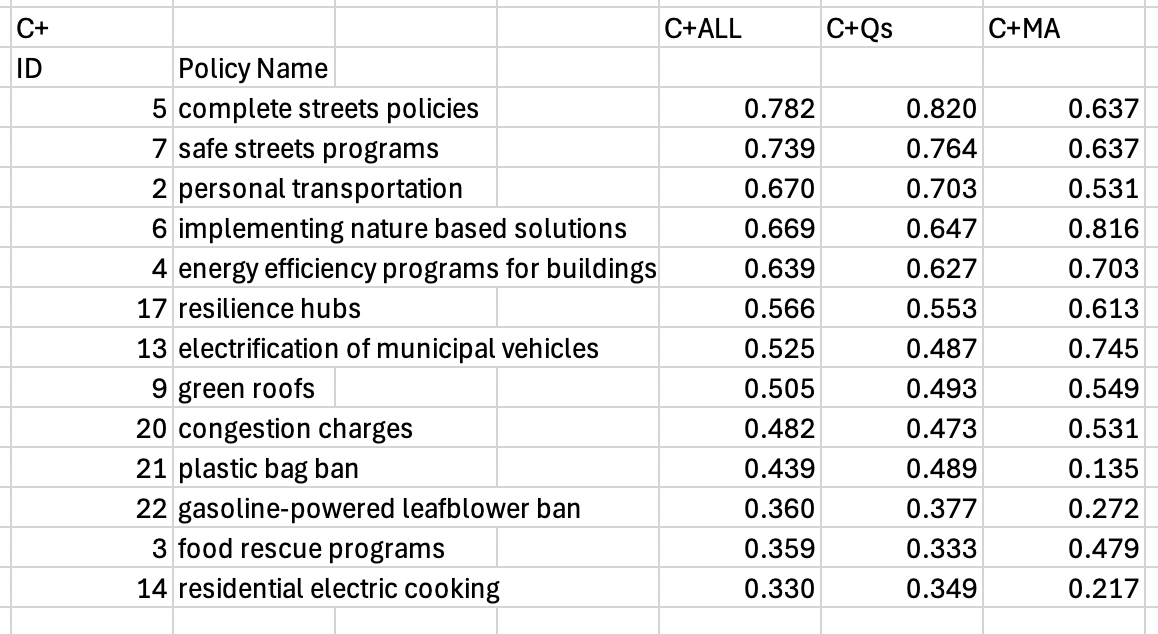}
    \caption{TOPSIS results. File: CurrentDatabase.xlsx, at \url{https://tinyurl.com/26azw7pp}.}
    \label{fig_topsis}
\end{figure}

We applied the TOPSIS MCDM method with equal weights on all 11 evaluation criteria (columns C--M in Figure \ref{fig_chatgpt_acs}). Figure \ref{fig_topsis} shows the results for three different subsets of the criteria. The column labeled C+All contains TOPSIS scores taking into account all 11 criteria. The column labeled C+Qs contains TOPSIS scores taking into account only the nine quality-of-life criteria (in columns E--M in Figure \ref{fig_chatgpt_acs}). Finally, the column labeled C+MA contains TOPSIS scores taking into account only the two criteria \emph{not} about quality-of-life (in columns C--D in Figure \ref{fig_chatgpt_acs}), viz., mitigation and adaptation.

Points arising:
\begin{enumerate}
    \item Figure \ref{fig_topsis} is sorted on column C+All in descending (reduced preference) order
    \item Column C+Qs is also in descending order, except for policies 6 and 2, indicating very strong agreement between the net evaluations of the quality-of-life criteria and these criteria augmented by the mitigation and adaptation criteria.
    \item The C+MA column is quite different from the other two columns in its ranking of the policies.  Policy 6, for example, is judged the best, while it is third in C+All and fourth in C+Qs. Policy 13 is even more divergent.
    \item Assuming that the ACS table scores are provisionally valid, we find that the rankings in Figure \ref{fig_topsis} are sufficiently sensible to serve as a point of departure for subsequent deliberations, accompanied by investigation and fine-tuning.
\end{enumerate}

\newpage
\section{Comparison with human scoring\label{sec_human_scoring}}

Recall that Table \ref{tbl_policy_set} (Appendix \ref{sec_master_reference_policy_list}, page \pageref{sec_master_reference_policy_list}) constitutes our master reference policy list for a series of MCDM studies on climate and sustainability policies.

We conducted an informed assessment exercise to score 21 of the 23 reference list policies on our nine quality-of-life evaluation criteria. In this exercise, members of the Climate Decisions Lab were given information about the several policies under consideration and asked to reflect on them. Individuals scored the policies on the nine quality-of-life criteria, using a 0--to--5, with 0 indicating irrelevance to or even negative impact on quality-of-life. Then the individual scores were averaged across multiple assessments. When multiple scorers were present, they had opportunity to discuss their scores and revise them.
Table \ref{tbl_p_table} contains the results of this exercise.

\begin{table}[h]
\centering
\begin{tabular}{lrrrrrrrrr}

\toprule
 ID & $Q_1$ & $Q_2$ & $Q_3$ & $Q_4$ & $Q_5$ & $Q_6$ & $Q_7$ & $Q_8$ & $Q_9$ \\
\midrule
0 & 2.5 & 1.5 & 3.5 & 1.0 & 0.0 & 0.0 & 1.0 & 1.0 & 2.0 \\
1 & 4.0 & 3.0 & 4.0 & 1.5 & 2.0 & 0.0 & 3.5 & 3.0 & 3.5 \\
2 & 4.5 & 4.0 & 5.0 & 3.0 & 4.0 & 3.5 & 4.0 & 3.5 & 4.5 \\
3 & 4.0 & 2.0 & 4.5 & 1.0 & 4.5 & 4.0 & 3.5 & 2.5 & 3.5 \\
4 & 3.0 & 4.0 & 3.0 & 1.0 & 1.0 & 0.0 & 3.0 & 5.0 & 4.0 \\
5 & 5.0 & 3.0 & 5.0 & 4.0 & 5.0 & 5.0 & 4.5 & 4.5 & 4.0 \\
6 & 3.0 & 2.0 & 4.0 & 4.0 & 3.0 & 1.0 & 5.0 & 3.5 & 4.0 \\
7 & 5.0 & 4.0 & 4.5 & 3.5 & 4.5 & 5.0 & 4.0 & 3.0 & 5.0 \\
8 & 5.0 & 4.0 & 3.5 & 5.0 & 3.0 & 3.0 & 3.0 & 2.5 & 5.0 \\
9 & 3.5 & 2.0 & 3.5 & 4.5 & 3.0 & 1.0 & 3.5 & 1.5 & 3.5 \\
10 & 4.0 & 5.0 & 4.5 & 3.0 & 4.0 & 5.0 & 3.0 & 3.0 & 4.5 \\
11 & 1.0 & 0.0 & 4.0 & 0.0 & 1.0 & 0.0 & 4.0 & 3.0 & 3.0 \\
12 & 4.0 & 4.0 & 3.5 & 3.0 & 4.0 & 5.0 & 3.0 & 4.5 & 4.0 \\
13 & 2.0 & 1.0 & 3.0 & 1.0 & 1.0 & 0.0 & 3.5 & 3.0 & 4.0 \\
14 & 4.0 & 4.2 & 2.0 & 4.0 & 2.0 & 2.0 & 2.1 & 2.3 & 3.0 \\
15 & 4.5 & 4.0 & 4.0 & 1.0 & 2.5 & 0.0 & 2.0 & 3.0 & 3.5 \\
16 & 5.0 & 3.0 & 5.0 & 3.5 & 4.5 & 4.5 & 3.5 & 4.0 & 4.5 \\
17 & 5.0 & 3.0 & 5.0 & 0.0 & 3.0 & 4.0 & 3.0 & 1.5 & 3.0 \\
18 & 5.0 & 4.0 & 5.0 & 3.0 & 3.0 & 3.0 & 1.5 & 2.5 & 4.0 \\
19 & 4.0 & 2.5 & 3.5 & 3.0 & 3.5 & 3.0 & 3.5 & 2.0 & 3.0 \\
20 & 3.0 & 0.0 & 3.0 & 2.0 & 2.0 & 2.0 & 2.0 & 3.0 & 2.5 \\
\bottomrule
\end{tabular}
\caption{$P$ (performance, aka ACS)  table  for well-being data. Row IDs are keyed to the policies in Table \ref{tbl_policy_set}. Column
headings are the evaluation criteria, discussed in Appendix \ref{sec_qol_indicators}, for well-being.
}
\label{tbl_p_table}
\end{table}

We then applied nine distinct rules or methods for decision making under uncertainty and/or MCDM to this table. The results are shown in Table \ref{tbl_e_decision_rules}.
\clearpage\newpage

\begin{table}[h]
\centering
\begin{tabular}{lrrrrrrrrr}
\toprule
ID & $D_1$ & $D_2$ & $D_3$ & $D_4$ &  $D_5$  & $D_6$ & $D_7$ & $D_8$ & $D_9$ \\
\midrule
0 & 4.5 & 2.0 & 4.5 & 1.5 & 1.0 & 6.6 & 12.5 & 1.5 & 1.0 \\
1 & 4.5 & 4.5 & 4.5 & 8.5 & 6.0 & 12.9 & 16.5 & 9.0 & 5.0 \\
2 & 18.5 & 16.0 & 18.5 & 17.0 & 17.0 & 18.9 & 4.0 & 20.0 & 17.0 \\
3 & 10.0 & 9.0 & 10.0 & 13.5 & 12.5 & 15.5 & 12.5 & 11.0 & 11.0 \\
4 & 4.5 & 16.0 & 4.5 & 8.5 & 9.0 & 12.6 & 20.5 & 5.0 & 7.0 \\
5 & 18.5 & 16.0 & 18.5 & 20.0 & 21.0 & 21.0 & 4.0 & 18.0 & 21.0 \\
6 & 10.0 & 16.0 & 10.0 & 13.5 & 12.5 & 15.5 & 16.5 & 13.0 & 13.0 \\
7 & 18.5 & 16.0 & 18.5 & 20.0 & 20.0 & 20.2 & 4.0 & 21.0 & 20.0 \\
8 & 15.0 & 16.0 & 15.0 & 13.5 & 15.0 & 17.8 & 9.0 & 17.0 & 15.0 \\
9 & 10.0 & 9.0 & 10.0 & 13.5 & 8.0 & 13.6 & 12.5 & 10.0 & 9.0 \\
10 & 18.5 & 16.0 & 18.5 & 17.0 & 18.0 & 18.9 & 4.0 & 19.0 & 16.0 \\
11 & 4.5 & 4.5 & 4.5 & 1.5 & 3.0 & 8.4 & 16.5 & 1.5 & 2.0 \\
12 & 18.5 & 16.0 & 18.5 & 17.0 & 16.0 & 18.4 & 4.0 & 15.0 & 18.0 \\
13 & 4.5 & 4.5 & 4.5 & 3.5 & 4.0 & 9.7 & 16.5 & 3.0 & 3.0 \\
14 & 13.5 & 7.0 & 13.5 & 5.0 & 5.0 & 13.4 & 8.0 & 6.0 & 10.0 \\
15 & 4.5 & 9.0 & 4.5 & 8.5 & 7.0 & 12.9 & 19.0 & 8.0 & 6.0 \\
16 & 18.5 & 16.0 & 18.5 & 20.0 & 19.0 & 19.7 & 4.0 & 16.0 & 19.0 \\
17 & 4.5 & 16.0 & 4.5 & 8.5 & 11.0 & 14.4 & 20.5 & 7.0 & 8.0 \\
18 & 12.0 & 16.0 & 12.0 & 8.5 & 14.0 & 16.3 & 12.5 & 14.0 & 14.0 \\
19 & 13.5 & 4.5 & 13.5 & 8.5 & 10.0 & 14.7 & 4.0 & 12.0 & 12.0 \\
20 & 4.5 & 1.0 & 4.5 & 3.5 & 2.0 & 10.2 & 10.0 & 4.0 & 4.0 \\
\bottomrule
\end{tabular}
\caption{$E$ (evaluations) table. Decision rule scores for the $P$ (performance or ACS) table. Data within columns are ranked (but based on scaled values in SAW, PROMETHEE, and TOPSIS). Produced by mcdm\_library\_climate\_wellbeing.ipynb.
$D_1$: maximin returns $D_2$: maximax returns $D_3$: minimax regret $D_4$: maximize medians $D_5$: lengths $D_6$: SAW $D_7$: Hurwicz $D_8$: PROMETHEE $D_9$: TOPSIS.}


\label{tbl_e_decision_rules}
\end{table}

Finally, we aggregated the scores (for the policies) in the rows of Table \ref{tbl_e_decision_rules} using three methods: the Borda count (row sums), the median, and the averaged rank median. The results are given in Table \ref{tbl_a_table}. The table is sorted by Borda count scores, but we note the strong agreement across the three measures.

\clearpage\newpage

\begin{table}[h]
\centering
\begin{tabular}{r|p{2in}rrr}
\toprule
 & & $V_1$ & $V_2$ & $V_3$ \\
 &  & Borda  & Simple  & Averaged  \\
 & Policy Name &  (Row Sums) &   Median &  
 Rank Median \\
\midrule
7 & Safe Streets programs & 158.21 & 20.00 & 20.00 \\
5 & Complete Streets Policy & 158.00 & 18.50 & 18.50 \\
16 & Active mobility pathways (alias micro-mobility) & 150.69 & 18.50 & 18.50 \\
2 & Personal transportation & 146.90 & 17.00 & 17.00 \\
10 & Public transit & 145.90 & 18.00 & 17.50 \\
12 & Transit Oriented Design & 141.38 & 17.00 & 16.00 \\
8 & Planting and nurturing of shade trees & 133.35 & 15.00 & 15.00 \\
6 & Nature based climate and sustainability solutions & 119.99 & 13.00 & 13.00 \\
18 & Facility amenities & 119.28 & 14.00 & 14.00 \\
3 & Food rescue programs & 104.99 & 11.00 & 11.00 \\
9 & Green roofs & 95.65 & 10.00 & 10.00 \\
17 & Resilience hubs & 94.44 & 8.50 & 8.50 \\
19 & Community gardens & 92.70 & 12.00 & 11.00 \\
4 & Energy efficiency programs for buildings & 87.60 & 8.50 & 7.00 \\
14 & Residential electric cooking & 81.44 & 8.00 & 8.00 \\
15 & Electrification of grounds care equipment & 79.36 & 8.00 & 7.00 \\
1 & Residential space heating and cooling with heat pumps & 71.36 & 6.00 & 6.00 \\
13 & Electrify Municipal Vehicles & 53.21 & 4.50 & 4.00 \\
11 & Municipal composting & 46.40 & 4.50 & 3.00 \\
20 & Congestion charges & 43.74 & 4.00 & 4.00 \\
0 & Residential hot water heating with heat pumps & 35.06 & 2.00 & 1.50 \\
\bottomrule
\end{tabular}
\caption{$A$ (aggregation) table. Three robust aggregations of the $E$ table.}
\label{tbl_a_table}
\end{table}

\section{Conclusion\label{sec_conclusion}}

\subsection{Comparing the GPT-4 and informed assessment results}

Comparing the Borda count results in Table \ref{tbl_a_table} with the TOPSIS results in Table \ref{fig_topsis} we see the following.
\begin{enumerate}
    \item There is good but hardly exact agreement between the GPT-4 rankings (under TOPSIS) and the Borda count rankings with TOPSIS as a component in essential agreement.
    \item Specifically, the GPT-4 ranking for the policies evaluated is 
    5, 7,  2, 6, 4, 17, 13, 9, 20, 21,  22, 3, 14.
    \item Informed assessment ranking of these policies (where evaluated) is 
    7, 5, 2, 6, 3, 9, 17, 4, 14, 13, 20. (Policies 21 and 22 were not evaluated by the informed assessment exercise. The TOPSIS alone ranking is 5, 7, 2, 6, 3, 14, 9, 17, 4, 20, 13; see Table \ref{tbl_e_decision_rules}.)
    \item Focusing on the eleven policies in common, the top four are 
    \begin{enumerate}
        \item GPT TOPSIS: 5, 7, 2, 6
        \item Informed assessment overall: 7, 5, 2, 6 
        \item Informed assessment TOPSIS: 5, 7, 2, 6
    \end{enumerate}
     
    \item The bottom four are 
    \begin{enumerate}
        \item GPT TOPSIS: 9, 20, 3, 14
        \item Informed assessment overall: 4, 14, 13, 20
        \item Informed assessment TOPSIS: 17, 4, 20, 13
    \end{enumerate}
     Again, decent agreement.
\end{enumerate}

In sum, recall our research question:

\begin{researchquestion}
    Can large language models and GPT in particular provide credible scores in ACS tables for climate and sustainability policies?
\end{researchquestion}
In this exploratory study we find that GPT-4 is in rough agreement with the policy rankings of informed assessment. Hence, we conclude (always provisionally and assuming a modest level of vetting) that GPT-4 can be used as a credible input, even starting point, for subsequent deliberation processes on climate and sustainability policies.

\subsection{Discussion}

Recall our goals and research question.

\setcounter{designgoal}{0}
\begin{designgoal}[tools]
    Construct tools to support deliberation on climate and sustainability policies. These tools may come in any of several forms, including knowledge of methods and practices, as well as databases, document corpora, and software.  The list of potential forms is open-ended. 
\end{designgoal}

\begin{designgoal}[starter MCDM models]
    Deliver to climate and sustainability policy stakeholders MCDM models (``starter MCDM models'') that are {\it prima facie}  acceptable, comprehensible, and suitable for follow on deliberation, revision, and evolution. These initial models should require minimal input from stakeholders, while affording revision in light of new information and stakeholder views when available.
\end{designgoal}

\begin{researchquestion}
    Can LLMs 
    provide credible evaluation scores, suitable for constructing starter MCDM models that support commencing deliberation regarding climate and sustainability policies?
\end{researchquestion}
As noted above, we find a positive answer to our research question. Using two approaches, one based on populating ACS tables using informed judgments and one using ChatGPT, we produced two {\it prima facie} MCDM starter models for preferential ranking of a number of policies. We found, moreover, that the two approaches are in reasonable agreement with each other. The upshot of this is that we arguably  have a Design Goal \ref{goal_mcdm_models} success example, and with that some positive steps toward meeting Design Goal \ref{design_goal_tools}. Put most briefly, we have 
demonstrated use of LLMs for construction of a starter MCDM model for policy evaluation, and we can expect such starter models will facilitate policy deliberation on climate change and sustainability. To this end we also have produced artifacts useful for deliberation and decision making:
\begin{enumerate}
    \item Goodness rankings of climate and sustainability policies, taking into account quality-of-life, climate mitigation, and climate adaptation.
    \item ACS tables of scores upon which the goodness rankings can be produced using various methods.
    \item A spreadsheet representation of the ACS tables and the TOPSIS MCDM method applied to them, affording thereby model analysis and revision in light of new information.
\end{enumerate}

\vskip 9 pt
\centerline{* * *}
\vskip 9 pt

We emphasize our \textit{\textbf{draft and revise}}  philosophy for modeling and deliberation (\S\ref{sec_background}). There are no real alternatives to it when, as in the present case of climate and sustainability policies, we have a dynamic environment in which the availability and credibility of relevant data is constantly in flux, is forever being contested, and is subject to continual need for revision in light of developments. Exacerbating the data problems, modeling and analyses based on them are concomitantly subject to revision. In short, we have the Nietzscheian condition of historicity and with it a fundamental barrier to definitional fixation.

How to proceed? The findings of this study suggest and support an approach that seeks a credible starting point (draft), followed by continual exploration, deliberation, and information collection, leading to revision and a new draft. The process is then repeated indefinitely. Decision are taken when necessary, based on the admittedly developing draft and assessment at the time of its reliability.

This process description should, we think, meet with general ascent for all inquiry. The upshot of this study concerns the specific, particular, and concrete matter of climate and sustainability policy making. We find that LLMs and informed assessment exercises applied in MCDM modeling plausibly serve as apt points of departure for {\it prima facie} bases for decision making and for follow on deliberation and investigation. These tools do in the present case and can more generally be used with confidence to produce first draft decision models and to support continued investigation.

\appendix

\newpage

\section{Supplementary Material\label{sec_supplementary_material}}

The {\it Climate and Sustainability Policies Sourcebook} and other documents referenced in this report can be found at \url{https://tinyurl.com/2yfe88st}. 

\section{Master reference policy list\label{sec_master_reference_policy_list}}


Table \ref{tbl_policy_set} constitutes our master reference policy list for a series of MCDM studies on climate and sustainability policies. This set was curated by the Climate Decisions Lab from the master database of climate and sustainability policies.
These are assembled in the spreadsheet file LocalPoliciesDatabase.xlsx, which may be found at \url{https://tinyurl.com/yyfydfnp}. The basis for this subjective curation (by many people) was the combined likelihood of being considered by municipalities, significance for mitigation of or adaptation to climate change, and effect on quality-of-life (alias well-being).

\begin{table}[h]
\centering
{\small
\begin{tabular}{rp{4in}}
\toprule
ID & Policy \\
\midrule
0 & Residential hot water heating with heat pumps \\
1 & Residential space heating and cooling with heat pumps \\ 
2 &  Personal transportation (EVs, scooters, e-bikes, etc.; supporting infrastructure such as charging stations, extra parking, etc.)
 \\ 
3 & Food rescue programs \\ 
4 & Energy efficiency programs for all kinds of buildings,
e.g., commercial, industrial, non-profit, and residential 
\\
5 & Complete Streets Policy  \\
6 &  Nature based climate and sustainability solutions  (Includes storm water
management, bio-swales, wetlands, green planted areas for
storm buffering, community forests (alias tiny forests), etc.)  \\
7 & Safe streets programs  (Includes safe routes to schools for
active mobility students.) \\ 
8 &  Planting and nurturing of shade trees  \\ 
9 &  Green roofs, for non-residential as well as residential buildings  \\
10 & Public transit \\ 
11 & Municipal composting   \\
12 &  Transit Oriented Design \\ 
13 &   Electrification of municipal vehicles  \\
14 & Residential electric cooking \\ 
15 & Electrification of grounds care equipment (grass cutting, hedge
trimming, leaf blowing, etc.) \\ 
16 & Active mobility pathways (alias micro-mobility). (For walking, biking, etc. Facilitating
both recreational use and utilitarian use, e.g., access to service
sites such as schools, shopping, restaurants, personal services.) \\ 
17 & Resilience hubs \\

18 & Facility amenities (bike racks at bus stops, shelters at bus stops
and other destinations, etc.) \\
19 &  Community gardens \\

20 & Congestion charges \\
21 & Plastic bag ban \\
22 & Gasoline-powered leaf blower ban \\
\bottomrule
\end{tabular}
}
\caption{Policy support consideration set, 23 in all. ID: index, unique identifier. Policy: name.}
\label{tbl_policy_set}
\end{table}

\section{Reference Policy Set for the Study\label{sec_ref_policies}} 


We essay to evaluate and compare policies \emph{available to municipalities} on the basis of their contributions to quality of life and well-being.  Our ultimate purpose is to identify for decision makers and the general public policy opportunities (and threats) to well-being. With such information to hand, policies that fail a standard cos-benefit analysis may on balance be deemed advisable. Moreover, policies that fare well on cost-benefit assessment, may be productively illuminated on the basis of quality of life.

We conducted Internet searches to assemble a list of more than 120 climate and sustainability policies that are in fact considered if not implemented by local governments (counties and municipalities in the USA). These are assembled in the spreadsheet file LocalPoliciesDatabase.xlsx, which may be found at \url{https://tinyurl.com/yyfydfnp}. Drawing on that database, the study team selected a subset of apparently promising policies to be targets of this study. The list is as follows:

\begin{enumerate}
\item personal transportation
\item food rescue programs
\item energy efficiency programs for buildings
\item complete streets policies
\item implementing nature based solutions
\item safe streets programs
\item green roofs
\item electrification of municipal vehicles
\item residential electric cooking
\item plastic bag ban
\item gasoline-powered leaf blower ban
\item resilience hubs
\item congestion charges
\end{enumerate}

\section{Reference Evaluation Criteria for the Study\label{sec_ref_criteria}}

These are (climate) mitigation and adaptation, (see ``Sustainability Policies Sourcebook'' in the Supplementary Materials), plus the nine quality-of-life criteria described in Appendix \ref{sec_qol_indicators}.

Mitigation prompt: 

In the context of climate change, mitigation refers to efforts to reduce or prevent the emission of greenhouse gases (GHGs) into the atmosphere, thereby minimizing the pace and impact of global warming. The goal of mitigation is to address the root causes of climate change by reducing emissions or enhancing ``sinks" (natural or artificial processes that absorb more carbon than they emit, such as forests, oceans, or carbon capture technologies).
Mitigation strategies can include:
Transitioning to Renewable Energy: Replacing fossil fuels with renewable energy sources like wind, solar, and hydroelectric power to reduce carbon emissions.
Energy Efficiency Improvements: Enhancing the efficiency of appliances, buildings, and industrial processes to use less energy.
Carbon Sequestration: Capturing and storing CO2 through techniques like reforestation, soil management, or carbon capture and storage (CCS) technologies.
Sustainable Transportation: Shifting to electric vehicles, public transportation, and other low-emission modes of transportation.
Changes in Agriculture and Land Use: Adopting sustainable farming practices and protecting or restoring forests and wetlands to maintain or increase natural carbon sinks.

Adaptation prompt:

Adaptation in the context of climate change refers to the process of adjusting systems (natural, social, or economic) to minimize the negative impacts of climate changes while maximizing potential benefits. It involves understanding the current and projected impacts of climate change and developing strategies to cope with them, ensuring resilience and sustainability.

\section{Quality-of-life evaluation criteria\label{sec_qol_indicators}}

As supplemental material for this study, the study team (alias Climate Decisions Lab at the University of Pennsylvania) developed a report entitled
{\it Climate and Sustainability Policies Sourcebook}. 
The {\it Sourcebook} 
may be found at: \url{https://tinyurl.com/2cw757ew}. We will draw upon it multiple times in this report. Typically, we excerpt the {\it sourcebook} for the sake of brevity in this report.

The study team undertook iterative literature reviews, brainstorming, and critical discussions to arrive at a list of nine quality of life criteria we found to be relevant to climate and sustainability policies. The list is as follows.


\subsection{Q-1: Health, safety, and hygiene.\label{bob}}

\subsubsection{Characterization}
\begin{itemize}
    \item [$Q_1$:] Health, safety, and hygiene. In the main, health considerations arise with (indoor or outdoor) air pollution, and amenities for exercise, active mobility, and recreation; safety mainly comes up with avoiding accidents, e.g., during transit; hygiene issues arise in multiple ways, perhaps most often in the design of parks and other public spaces. Both physical and mental health fall under this criterion.
\end{itemize}

\subsubsection{Version for prompting\label{q1_evaluation_prompt}} 
    \begin{quote}
Health, safety, and hygiene. In the main, health considerations arise with (indoor or outdoor) air pollution, and amenities for exercise, active mobility, and recreation; safety mainly comes up with avoiding accidents, e.g., during transit; hygiene issues arise in multiple ways, perhaps most often in the design of parks and other public spaces. Both physical and mental health fall under this criterion.
\end{quote}

\subsection{Q-2: Time, attention and convenience.}

\subsubsection{Characterization}
    \begin{itemize} 
    \item [$Q_2$:] Time, attention and convenience. Policies that economize on time or  relieve people of having to maintain an item or that in some way relieve people of inconveniences score highly. Policies that may increase discretionary time for people should score highly. Conversely, policies that may reduce discretionary time should receive very low scores \citep{goodin_discretionary_2008}. Discretionary time may be characterized as follows:
    \begin{quote}
        The time beyond that necessary to attend to necessary functions is yours to use as you please. That is what we will call `discretionary time'. That is how much `temporal autonomy' you possess. \citep[page 5]{goodin_discretionary_2008}
    \end{quote}
    
    To illustrate, some respondents find electric vehicles very attractive on this criterion, typically because home charging and reduced maintenance for electric vehicles frees up time that had to be spent on internal combustion engine (ICE) vehicles for refueling and scheduled maintenance. Other respondents report to the contrary. Typically this is because they lack access to convenient charging and so would need to commit more time to refueling an electric vehicle than an ICE vehicle. 

    To take another example, a policy that reduced commuting times would perforce be increasing discretionary time for the commuters affected.
    \end{itemize}
    \subsubsection{Version for prompting}

    \begin{quote}
        Time, attention and convenience. Policies that economize on time or  relieve people of having to maintain an item or that in some way relieve people of inconveniences score highly. Policies that may increase  time for people to use as they please should score highly. Conversely, policies that may reduce freely disposable time should receive very low scores. 
    \end{quote}

\subsection{Q-3: Moral considerations.}

\subsubsection{Characterization}
    \begin{itemize}
    \item [$Q_3$:] Moral considerations. This dimension of evaluation is meant to record broadly moral or ethical aspects of the policy in question. These ethical considerations may constitute a reason why the policy should be implemented, or they may enter as a consequence of implementing the policy. The single criterion links with the flourishing moral foundations literature (empirical study of ethics and morality), compressing several dimensions into one. See \cite{adger_moral_2017,haidt_righteous_2012,culiberg_role_2023}. We were surprised when these considerations arose spontaneously and fairly frequently in a number of semi-structured interviews we conducted, both in Taiwan (described above) and in the U.S. For this reason, we include moral considerations in our well-being criteria.
\end{itemize}

\subsubsection{Version for prompting}
    \begin{quote}
         Moral considerations. This dimension of evaluation is meant to record broadly moral or ethical aspects of the policy in question. These ethical considerations may constitute a reason why the policy should be implemented, or they may enter as a consequence of implementing the policy. The single criterion links with the flourishing moral foundations literature (empirical study of ethics and morality), compressing several dimensions into one.
    \end{quote}

    

\subsection{Q-4: Aesthetics}

\subsubsection{Characterization}

\begin{itemize}
    \item [$Q_4$:] Aesthetics. Policies score favorably  if they result in improvements in how aesthetically pleasing the local environs are. Aesthetic concerns were often mentioned in interviews and the online survey comments.
    \end{itemize}

    \subsubsection{Version for prompting}

    \begin{quote}
        Aesthetics. Policies should be scored favorably  if they result in improvements in how aesthetically pleasing the local environs are. 
    \end{quote}

\subsection{$Q_5$: Community strengthening.}

\subsubsection{Characterization}
    \begin{itemize}
    \item [$Q_5$:] Community strengthening. Policies score favorably if their implementation would tend to build stronger, more cohesive communities, fostering a sense of togetherness. This could be expected to happen, for example, with policies that     
    afford  activities and interactions in and among the public. We are assuming the policy is well or at least adequately implemented.
    \end{itemize}
    \subsubsection{Version for prompting}

    \begin{quote}
        Community strengthening. Policies score favorably if their implementation would tend to build stronger, more cohesive communities, fostering a sense of togetherness. This could be expected to happen, for example, with policies that  
    afford  activities and interactions in and among the public. It should be assumed that the policy is well or at least adequately implemented.
    \end{quote}

\subsection{$Q_6$: Access to service points}

\subsubsection{Characterization}
\begin{itemize}
    \item [$Q_6$:] Access to service points.  Service points include just about any location that would be a destination for a trip, such as grocery stores, schools, places of work, pharmacies, libraries, and retail outlets generally. A policy scores well to the extent that it improves ease of access to services or is itself a service to which access is valuable.
    \end{itemize}
\subsubsection{Version for prompting}

\begin{quote}
  Access to service points.  Service points include  any location that would be a destination for a trip, such as grocery stores, schools, places of work, pharmacies, libraries, and retail outlets. A policy scores well to the extent that it improves ease of access to services or is itself a service to which access is valuable.
   \end{quote} 

\subsection{$Q_7$: Affordance of capabilities}

\subsubsection{Characterization}

    \begin{itemize}
    \item [$Q_7$:] Affordance of capabilities. Policies that score well make positive contributions to the exercise and development of human capabilities, including physical (exercise), mental, and social capabilities. Prototypically, this can be done through education, learning, knowledge acquisition, and enriching, uplifting, broadly educative experiences.
    
\end{itemize}
\subsubsection{Version for prompting}

\begin{quote}
    Affordance of capabilities. Policies that score well make positive contributions to the exercise and development of human capabilities, including physical (such as exercise and athletics), mental, and social capabilities. This is typically done through education, learning, knowledge acquisition, and generally through enriching, uplifting, broadly educative experiences.
\end{quote}

\subsection{$Q_8$: Economy}

\subsubsection{Characterization}

    \begin{itemize}
    \item [$Q_8$:] Economy.  Policies that score well on economy make positive contributions to saving  money, or they stimulate employment, or they attract investment, or they tend to increase tax revenues (without raising tax rates), and so on.
\end{itemize}

\subsubsection{Version for prompting}

\begin{quote}
    Economy.  Policies that score well on economy make positive contributions to saving  money, or they stimulate employment, or they attract investment, or they tend to increase tax revenues (without raising tax rates), and so on.
\end{quote}


\subsection{$Q_9$: Improving and creating  destinations.}

\subsubsection{Characterization}

\begin{itemize}
        \item [$Q_9$:] Improving and creating available destinations. $Q_6$ is about improving access to existing destinations. $Q_9$ is about improving the collection of destinations worth accessing. Policies that score well make positive contributions to creating or improving destinations or to making certain destinations more valuable or available to the public. For example, a nature-based storm water management policy may lead to the creation or improvement  of a natural setting that would be suitable for recreational purposes at least much of the time. The destinations attribute is distinct from the accessibility attribute (access to service points, $Q_6$). Accessibility is about easing travel to a fixed destination. The destinations attribute is about making it more attractive to bother traveling to a place.
\end{itemize}

\subsubsection{Version for prompting}

\begin{quote}
    Improving and creating available destinations is about enhancing the collection of destinations worth accessing. Policies that score well make positive contributions to creating or improving destinations or to making certain destinations more valuable or available to the public. For example, a nature-based storm water management policy may lead to the creation or improvement  of a natural setting that would be suitable for recreational purposes at least much of the time. To take another example, measures to improve safety and pleasantness of pedestrian access may improve the attractiveness of a retail corridor.
\end{quote}


\bibliographystyle{apalike}
\bibliography{ClimateDecisionsLab}
\end{document}